\documentclass[a4,11pt]{article}                     
\usepackage{graphicx}
\usepackage{pifont}
\usepackage{geometry}
\usepackage{varioref}
\usepackage{natbib}
\usepackage{mathrsfs}
\usepackage{amsmath}
\usepackage{amsfonts}
\usepackage{amssymb}
\usepackage{pst-all}
\usepackage{comment}
\usepackage{url}

\newtheorem{defn}{Definition}
\newtheorem{conject}{Conjecture}

\begin{document} 
\title{Causation, Measurement Relevance and No-conspiracy in EPR\thanks{Support from the Spanish Ministry of Science and Innovation (FFI2008-06418-C01-03), and the Education Department of Madrid's Regional Government (S2007-HUM/0501) is gratefully acknowledged.}} 
\date{~}
\author{I{\~n}aki San Pedro\thanks{Universidad Complutense de Madrid}}
\maketitle

\begin{abstract}
In this paper I assess the adequacy of \emph{no-conspiracy} conditions employed in the usual derivations of the Bell inequality in the context of EPR correlations. First, I look at the EPR correlations from a purely phenomenological point of view and claim that common cause explanations of these cannot be ruled out. I argue that an appropriate common cause explanation requires that \emph{no-conspiracy} conditions are re-interpreted as mere \emph{common cause-measurement independence} conditions. In the right circumstances then, violations of measurement independence need not entail any kind of conspiracy (nor backwards in time causation). To the contrary, if measurement operations in the EPR context are taken to be causally relevant in a specific way to the experiment outcomes, their explicit causal role provides the grounds for a common cause explanation of the corresponding correlations.      
\end{abstract}                   

\section{Introduction}
\label{sec:intro}
This paper aims to provide an analysis of some our most robust intuitions about causation, with the ultimate purpose of testing them in the context of quantum mechanics. Philosophers wanting to understand causation need to face, for instance, the adequacy of the diverse methods of causal inference available. Reichenbach's Principle of the Common Cause (RPCC) is one such method, and it is precisely in the quantum mechanical context that RPCC faces one of its most significant threats. I shall not discuss the specific problems related to the philosophical status of RPCC ---these are diverse and interesting in themselves even if quantum mechanics is not brought into the picture. RPCC will simply be assumed to hold, for the sake of the argument, and the causal picture that it purports tested in the quantum context.

The received view, though, takes it that (Reichenbachian) common cause accounts of EPR correlations are to be ruled out~\citep{butterfield1989, butterfield2007, cartwright1987, grasshoff2005, vanfraassen1982a}. The standard argument views causes as hidden variables onto which several constraints are set. These constraints are intended to reflect standard requirements  typical of any physical system, including temporal order of causal relations and considerations about locality. As a result, some Bell-type inequality is derived. The strength of such arguments relies, as it stands, on the plausibility of the conditions imposed on the common causes.

\emph{No-conspiracy} is one such condition present in the usual derivations of the Bell inequalities. The motivation behind it is that the postulated common causes be independent of the measurement settings. Violations of such independence are standardly interpreted as to entail certain strange `conspiratorial behaviour', unless super-determinism or backwards in time causation is brought into the picture~\citep{berkovitz2002, price1994, szabo2000, szabo2008}.

I shall challenge this standard reading of no-conspiracy-type conditions. I will do this in two steps. I will first look at the very formal structure of such conditions, while remaining neutral as to whether their violation entail any kind of conspiracy. I will then look at the EPR correlations from a purely phenomenological point of view and assess the role of measurement in the EPR experiment itself. This will suggest a re-interpretation of no-conspiracy-type conditions, under which violations of these can be accommodated with no conspiratorial implications attached to them, nor backwards in time causation. I claim that this new view also provides the grounds for a potential common cause model of EPR correlations. Although the model is not to be developed in full here, I shall sketch what its main features could be and discuss its potential consequences as regards non-locality.

The structure of the paper is as follows. Section~\ref{sec:eprrpcc} provides the necessary background on the EPR experiment and Reichenbach's Principle of the Common Cause respectively. In Section~\ref{sec:structure} the formal structure of the problem is stated and the idea of \emph{no-conspiracy} introduced. Section~\ref{sec:phenomenological} looks at the EPR experiment again but this time from a purely phenomenological point of view, and examines the role of measurement in it. This analysis will eventually identify the main features of a potential common cause model of EPR, which is developed (although only partially) in Section~\ref{sec:themodel}. The implications of this partial model as regards locality are discussed in Section~\ref{sec:mdependence}, together with some open questions for further investigation.

\section{EPR correlations and Common Causes}
\label{sec:eprrpcc}
Consider the Bohm version of the EPR experiment. Two entangled electrons are emitted from a source in opposite directions. The spin component of each of the electrons can be later detected (measured) in three different directions $\theta_i ~(i=1,2,3)$ after having passed through an inhomogeneous magnetic field. It is assumed that the state of the entangled electron pair at the source is the spin \emph{singlet} state: 
\[
\Psi_s = \frac{1}{\sqrt{2}} (\psi^{+}_{L} \otimes \psi^{-}_{R} - \psi^{-}_{L} \otimes \psi^{+}_{R}),
\]
\noindent where $\psi^{+}_{L}$ ($\psi^{+}_{R}$)  denotes the \emph{up} eigenvector of the spin component of the particle in the left (right) wing of the experiment along a given direction. Similarly for $\psi^{-}_{L}$ ($\psi^{-}_{R}$) and the \emph{down} eigenvector of the spin component of the corresponding particle.

Let us write $L_{i}$ and $R_{j}$ for the different measurement settings in each wing of the experiment. Also, we shall denote by $a$ and $b$ the value of the spin variable of each electron in a given measurement direction which, for the singlet state, can be either `spin-up' ($+$) or `spin-down' ($-$), with probability $\frac{1}{2}$. Measurement outcome events on each particle along each of the three different measurement directions will be written as $L^{a}_i$ and $R^{b}_{j}$, with $a,b = +,-$ and $i,j = 1,2,3$.

It is assumed that both measurements and outcome events at each wing of the experiment, $L^{a}_{i}$ and $R_{j}^{b}$, are space-like separated. Quantum mechanical probabilities for the possible outcomes of the experiment are given by the trace $Tr (\hat{W} \hat{A})$, where $\hat{W}$ is the density operator corresponding to the singlet state $\Psi_s$ and $\hat{A}$ is the projector associated to a particular property of the system ---say, for instance, `spin-up' along direction $i$.  

Quantum mechanics predicts correlations between the outcomes measured in the three possible directions in both wings. We would like to know whether these correlations are the result of underlying causal processes. At this point already, the idea of common cause seems rather appealing. For the space-like separation of the outcomes at each wing of the experiment seems to rule out, at least in principle, direct causal interactions between these. And this is in fact the typical situation common cause explanations are suitable for. 

The principle of the common cause was first introduced by \cite{reichenbach1956}. It states, in short, that there are \emph{no correlations without causal explanation} (either in terms of direct causal interactions or by means of a common cause).

The idea of correlation is defined within the framework of classical Kolmogorovian probability spaces as:\footnote{This definition is of positive correlation. A completely symmetrical definition may be given for negative correlations. Distinguishing between positive and negative correlations will not be important for the argument here. Thus, if not stated otherwise, positive correlations will be assumed throughout the paper.}

\begin{defn}
\label{defn:corr}
Let $(\mathcal{S},p)$ be a classical probability measure space with Boolean algebra $\mathcal{S}$ representing the set of random events and with the probability measure $p$ defined on $\mathcal{S}$. If $A,~ B \in \mathcal{S}$ are such that
\begin{equation}
\label{eqn:def_corr}
Corr_{p}(A,B) = p(A \land B) - p(A) \cdot p(B) > 0,
\end{equation}
then the event types $A$ and $B$ are said to be (positively) correlated.
\end{defn}

The principle of the common cause can then be formalised, following \citeauthor{reichenbach1956}'s own ideas, as follows:

\begin{defn}[RPCC]
\label{def:rpcc}
For any two (positively) correlated event types $A$ and $B$ ($Corr_p(A, B) > 0$), if $A$ is not a cause of $B$ and neither $B$ is a cause of $A$,  there exists a common cause $C$ of $A$ and $B$  such that the following independent conditions hold:
\begin{align}
\label{eq:rcc1}
p(A \land B \vert C) & = p(A \vert C) \cdot p(B \vert C) \\
\label{eq:rcc2}
p(A \land B \vert \neg C) & = p(A \vert \neg C) \cdot p(B \vert
\neg C) \\ 
\label{eq:rcc3}
p(A \vert C) & > p(A \vert \neg C) \\
\label{eq:rcc4}
p(B \vert C) & > p(B \vert \neg C) 
\end{align}

\noindent where $p(A \vert B) = p(A \land B)/p(B)$ denotes the probability of $A$ conditional on $B$ and it is assumed that none of the probabilities $p(X)~ (X = A, B, C, \neg C)$ is equal to zero.
\end{defn}

The philosophical status of RPCC is not completely settled. There is no need to address here the various philosophical problems that RPCC faces, though, and I shall assume in what follows that the principle, as stated above, holds.\footnote{See \cite{suarez2007} and \cite{sanpedro-suarez2009} for a detailed assessment of the status of RPCC.} I shall refer to common causes postulated by Definition~\ref{def:rpcc} irrespectively as \emph{Reichenbachian common causes}, \emph{screening-off common causes} or simply \emph{common causes}. Endorsing RPCC may be motivated by at least two reasons. First, note that for Reichenbach the role of the principle is mainly explanatory~\citep[p.~159]{reichenbach1956}. The\label{remark:explanation} explanatory power of screening-off common causes may thus be taken as a good methodological reason to support the adequacy of RPCC for the inference of causal relations from probabilistic facts, even if it can be patently shown not to hold as a necessary nor as a sufficient condition for common causes.

Second, recent results show that, at least formally, it is always possible to provide a Reichenbachian common cause for any given correlation. These results build on the intuition that any probability space $\mathcal S$ containing a set of correlations and which does not include (Reichenbachian) common causes of these, may be extended in such a way that the new probability space $\mathcal{S}'$ \emph{does} include (Reichenbachian) common causes for each of the original correlations. This is formalised in so-called \emph{extendability} and \emph{common cause completability} theorems~\citep{hofer-szabo1999,hofer-szabo2000a}.

\emph{Common cause completability} thus constitutes a very powerful tool if we are to provide common cause explanations of generic correlations. The whole program faces however its own problems, especially when it comes to the physical interpretation of either the enlarged probability space $\mathcal{S}'$ or the new common causes contained in it.\footnote{I point the reader to \cite{sanpedro-suarez2009} for a recent assessment the significance of common cause completability, possible criticisms to it and the strategies to avoid these.} In particular, it seems a fair criticism to the program to claim that common cause completability is merely a formal device, which is likely to lack physical meaning in many (perhaps too many) cases. To be more precise, what \citeauthor{hofer-szabo1999} show is that, for any given correlation, there exist formal objects (events on the algebra) which conform to Definition~\ref{def:rpcc}. These we call (Reichenbachian) common causes, as pointed out, even if they may well lack physical (causal) interpretation. Thus, what \emph{common cause completability} provides, rather than a proof that there exist common causes (with causal/physical meaning this time) for any correlation, is a proof that these cannot be ruled out. All in all, and despite the limitations of the formal results by \citeauthor{hofer-szabo1999}, the idea of \emph{common cause completability} encodes a bunch of good intuitions as to what we should require or expect when looking for common cause explanations. \emph{Common cause completability} can thus be taken as a good heuristic or methodological tool when aiming at explaining correlations in terms of common causes. It is with this remarks in mind that \emph{common cause completability} is invoked here.   

There is a further important remark as regards \emph{common cause completability}. It has to do with the distinction between so-called \emph{individual}-common causes and \emph{common}-common causes. We call \emph{individual}-common causes those events satisfying~\eqref{eq:rcc1}-\eqref{eq:rcc4} for a single (individual) correlation \emph{only}. \emph{Common}-common causes, on the other hand, will satisfy~\eqref{eq:rcc1}-\eqref{eq:rcc4} for a set of two or more correlations. This is an important distinction since, while \emph{individual}-common cause completability holds in general for every classical probability space $\mathcal S$, this is not the case for \emph{common}-common causes. In other words, while \emph{common cause completability} guarantees that \emph{individual}-common causes exist (at least formally) for any given correlation, this is not true in general for \emph{common}-common causes~\citep{hofer-szabo2002}. In what follows and if not stated otherwise, when common cause is written it will mean \emph{individual}-common cause.

\section{Structure of the problem}
\label{sec:structure}
We now seem to be in a position to attempt at an answer to the question as to whether an explanation of the EPR correlations can be given in terms of common causes. In order to proceed it is convenient to interpret quantum probabilities, i.e~trace-like quantities $Tr (\hat{W} \hat{A})$, as classical conditional probabilities ---conditional on measurement operations, that is.\footnote{This is indeed standard and, as pointed out by~\cite[p.~4]{szabo2000}, taking quantum probabilities as classical conditional probabilities is crucial for the whole issue regarding Reichenbachian common cause explanations of EPR correlations to be meaningful. As an alternative we would need to redefine RPCC to fit the non-classical probabilistic framework of quantum mechanics. This option is explored for instance in~\cite{henson2005}.} Thus the `quantum probability' that the particle in the left wing of the experiment, for instance, is measured with spin-up will be given by $p(L_i^+ \vert L_i)$.

Now if we recall \emph{common cause completability}, the answer to our question seems, at least in principle, a positive one. The issue is not so simple, though. For we cannot overlook the fact that if our postulated common causes are to be physically sensible at all, they will need to fulfil certain requirements ---besides those in their definition, i.e.~equations~\eqref{eq:rcc1}-\eqref{eq:rcc4}. The problem becomes thus more complex since we first need to identify and characterise these extra (physically sensible) requirements Reichenbachian common causes must conform to.

Such conditions typically include those intended to capture in some sense or another the idea of physical locality ---so as to avoid conflict with special relativity.\footnote{What it is exactly meant by `physical locality' and whether such concept may be appropriately captured in terms of probabilistic relations, though, is far from settled. I shall not discuss these issues in detail here and just point the reader to \cite{butterfield2007, fine1981, fine1986, maudlin1994, wessels1985} or \cite{suarez2000} for further reference.} But requiring constraints of this type basically leads to the Bell inequalities, and the usual move is then to give up the idea of Reichenbachian common cause. It is not completely clear however whether this needs to be the case. Indeed we may want to insist in providing a Reichenbachian common cause explanation of the correlations while avoiding the charge of Bell's theorem. The strategy would then be to define a set of conditions that, along with the idea of Reichenbachian common cause, provides a satisfactory explanation of the correlations without being committed to the Bell inequalities. Ideally we would like to make sure that such restrictions do not entail the idea of \emph{common}-common cause either. For, as I have pointed out it is not likely that \emph{common}-common cause explanations exist for any set of correlations in general. 

One of the most remarkable attempts to provide a common cause model for EPR following this strategy is \cite{szabo2000}.  \citeauthor{szabo2000}'s model is explicitly construed by recursively applying the idea of \emph{common cause completability}. It turns out however that the model features unwanted dependencies between certain combinations of the postulated common causes and the measurement setting events. These dependencies are usually interpreted as conspiratorial.

Avoiding such conspiratorial features is indeed the role of so-called \emph{no-conspiracy} conditions which are also standard in the derivations of the Bell inequalities. Roughly, \emph{no-conspiracy} conditions require statistical independence between the common cause $C$ and the measurement settings:
\begin{align}
p(L_i \land C) & = p(L_i) \cdot p(C),\label{eq:simple-no-cons1} \\
p(R_j \land C) & = p(R_j) \cdot p(C).\label{eq:simple-no-cons2}
\end{align}

The intuition, clearly, is that the postulated common causes must not have an influence, or be otherwise influenced, by the choices of measurement settings. In the usual interpretation, the existence of such influences ---i.e.~violations of the equations~\eqref{eq:simple-no-cons1} and~\eqref{eq:simple-no-cons2}--- would be conspiratorial in the sense that the taking place of the common cause would somehow `force' or determine to some extent the presumably free independent decisions of the experimenter about measurement.

Equations~\eqref{eq:simple-no-cons1} and~\eqref{eq:simple-no-cons2} may be regarded as `simple' \emph{no-conspiracy} conditions, in that they involve only one common cause event at a time. More complex conditions may involve combinations (both conjunctions and disjunctions) of the different postulated \emph{individual}-common causes, e.g.~$p(L_i \land C_{ij}^{ab} \land C_{i'j'}^{a'b'})$ ---these complex no-conspiracy conditions are actually what \cite{szabo2000}'s model fails to satisfy. We need not discuss the later case here. For I will be arguing that violations of the `simple' case, i.e. of equations~\eqref{eq:simple-no-cons1} and~\eqref{eq:simple-no-cons2}, are consistent with a plausible common cause explanation of EPR correlations. And, of course, violations of the `simple' no-conspiracy conditions also entail violations of the more complex relations.

The significance of \emph{no-conspiracy} assumptions relies heavily on another underlying implicit assumption. This is the assumption that the postulated common causes take place \emph{before} the apparatus are set for measurement, and therefore also \emph{before} (always in the rest-frame of the laboratory) measurement is performed. And it is only by assuming this specific time-order framework that one can make sense of the standard interpretation involving claims about world conspiracies and free will (or the lack of it).\footnote{In fact, I think this is why discussions about free will and backwards in time causation are so much entangled. For those defending backwards causation still assume (temporal) priority of common causes in relation to measurement operations. (Only, in those cases, time order and causal order are not assumed to coincide.) See, for instance, \cite{berkovitz2002} or \cite{price1994}.} This is a crucial assumption but in my view it is not completely warranted. Indeed, I shall challenge it and suggest that violations of~\eqref{eq:simple-no-cons1} and~\eqref{eq:simple-no-cons2} do not necessarily entail that there being world conspiracies. In order to keep neutral as regards the possible conspiratorial implications of the violation of expressions~\eqref{eq:simple-no-cons1} and~\eqref{eq:simple-no-cons2}, I shall refer to them as \emph{measurement independence} conditions.

\section{EPR from a phenomenological point of view}
\label{sec:phenomenological}
I would now like to tackle the issue from a different point of view. More specifically, I shall not look at the EPR correlations as `quantum correlations' ---i.e.~as correlations defined between conditional probabilities, as suggested in \cite{szabo2000}---, and then postulate the corresponding \emph{individual}-common causes. Instead, I shall look at the EPR correlations from a completely classical perspective first, where entirely \emph{classical} common causes can be postulated for them. Only later I shall provide these common causes with an appropriate quantum mechanical interpretation. 

Suppose then we are provided with two lab record notebooks, each containing the results of the corresponding experiments at each wing of the EPR-Bohm set-up. We are to look for correlation patterns among our raw phenomenological data. The data analysis is at a completely classical level, and we can therefore set aside any considerations related to quantum mechanics at this point. (The data, in other words, is to be treated classically, regardless of its quantum origin.) Suppose for the sake of the argument that, faced with the correlated data, we can rule out with certainty direct causal connections between the correlated variables. This points then to the possibility of explaining the correlations in terms of common causes. In fact, since the correlations are defined over a completely classical framework, i.e.~among events that belong to a Boolean algebra, we may want to take advantage of \emph{common cause completability} and set for a common cause explanation.\footnote{The actual process of finding screening-off events (potential common causes) may consist in recursively applying \emph{common cause completability}, as in \cite{szabo2000}. We should keep in mind however the various limitations of this approach. (See my remarks in Section~\ref{sec:eprrpcc} for details).} What would the structure of such \emph{classically} postulated common causes be? 

The first thing to note is that the postulated common causes are Reichenbachian \emph{individual}-common causes. (This is so by construction). That is to say, they are common causes which screen-off one, and only one, of the correlations. Consequently, they will have a label which identifies the specific correlation, i.e.~$a, b= +,-$. On the other hand, each correlation found among our record notebooks data sets refers explicitly to a particular `joint experiment', which corresponds to the specific measurement operations carried out on both wings of the EPR-Bohm set-up. The postulated common cause will the be labelled accordingly as well, i.e.~$i,j=1,2,3$. Let us denote such an event as $\mathsf{C_{ij}^{ab}}$.\footnote{I am following here basically the same notation to that in the rest of the paper. Only I shall denote the common cause of the classical correlations with a \textsf{\small{serif}}~font~$\mathsf{C_{ij}^{ab}}$ instead of the usual \textit{italic}~$C_{ij}^{ab}$ used in standard treatments of the problem in order to stress their exclusively classical origin.} We will have then:
\begin{align}
p(L_i^a \land R_j^b \vert \mathsf{C_{ij}^{ab}}) & = p(L_i^a \vert \mathsf{C_{ij}^{ab}}) \cdot p(R_j^b \vert \mathsf{C_{ij}^{ab}}), \label{eq:model_RPCC_01}\\
p(L_i^a \land R_j^b \vert \lnot \mathsf{C_{ij}^{ab}}) & = p(L_i^a \vert \lnot \mathsf{C_{ij}^{ab}}) \cdot p(R_j^b \vert \lnot \mathsf{C_{ij}^{ab}}), \label{eq:model_RPCC_02}\\
p(L_i^a \vert \mathsf{C_{ij}^{ab}}) & > p(L_i^a \vert \lnot \mathsf{C_{ij}^{ab}}),\label{eq:model_RPCC_03}\\
p(R_j^b \vert \mathsf{C_{ij}^{ab}}) & > p(R_j^b \vert \lnot \mathsf{C_{ij}^{ab}}).\label{eq:model_RPCC_04}
\end{align}

The fact that both the correlations and thus the postulated common causes refer to specific `joint experiments' ---which are defined, as pointed out, by specifying the actual measurement operations performed on \emph{both} wings of the EPR-Bohm set-up--- may be interpreted as to suggest that the $\mathsf{C_{ij}^{ab}}$ events have a causal component due to \emph{both} the EPR measurement operations. That is, the postulated common causes can be thought as to \emph{contain} or \emph{include} somehow information about the measurement operations. In other words, \emph{measurement operations are causally relevant to the common causes} in this picture. This is indeed a remarkable feature of the postulated common causes, although far from uncontroversial.\footnote{In the context of EPR correlations it is indeed standard to assume measurement operations not to be causally relevant for common causes ---this is in fact why `no-conspiracy'-type conditions are required. Note moreover that the expression `measurement operations' here include not only the experimenter's act of setting-up the apparatus but also the actual interaction between the particle and the apparatus when measurement is preformed.} 

Note however that measurement operations need not be taken as the only relevant causal factors of the common causes. In particular, there seem to be good intuitive grounds to say that the postulated common cause events will include some characteristic feature of the system at hand. It is very often assumed that the spin-singlet state itself ---or perhaps some factor closely related to it--- is causally relevant to the EPR outcomes.\footnote{This is indeed commonplace~\citep{cartwright1987, cartwright-jones1991, cartwright-chang1993, healey1992, price1994, suarez2007, vanfraassen1982a}.} Finally, the postulated common causes will not necessarily be deterministic, i.e.~their occurrence will not necessarily entail that the corresponding outcomes will occur with certainty.

\section{A (partial) common cause model for EPR correlations}
\label{sec:themodel}
The above can be expressed formally, in terms of algebra events and probabilistic relations. These shall provide the grounds for a potential common cause model of EPR. I shall not develop such a model in full here but only sketch its most salient features and discuss their consequences. In this sense, what follows is to be seen as a \emph{partial} or \emph{incomplete} common cause model of EPR. I shall nevertheless refer to it as `model' throughout.

\subsection{\emph{Individual}-common causes and outcome independence}
\label{ssec:ccstructure}
Recall first that common causes were each postulated to screen-off a single correlation ---they are what I called \emph{individual}-common causes. I expressed that by means of the screening-off conditions~\eqref{eq:model_RPCC_01}-\eqref{eq:model_RPCC_02}.
 
Now the causal role of measurement can be explicitly accounted for if we think of the postulated \emph{individual}-common causes $\mathsf{C_{ij}^{ab}}$ as events of the form
\begin{align}
\label{eq:c_in_lirilambda}
\mathsf{C_{ij}^{ab}} & \subset L_i \land R_j \land \Lambda,
\end{align}

\noindent where $L_i$ and $R_j$ are the specific measurement operations and $\Lambda$ is a futher causal factor.\footnote{Why $\mathsf{C_{ij}^{ab}}$ is defined as a subset of the conjunction $L_i \land R_j \land \Lambda$, and not as identical to it, will become clear in a moment.} Typically, $\Lambda$ will most probably be associated to the singlet state $\Psi_s$ itself, and perhaps to some other relevant causal factors prior to the preparation of the entangled system.\footnote{This characterisation of $\Lambda$ reminds somehow to the kind of events Cartwright considers the right common cause events for EPR. That the similarity is quite so it will become clear in a moment, since I will be requiring (or at least allowing) that the $\Lambda$ be non-screening-off events, just like in Cartwright's common cause account of EPR~\citep{cartwright1987, cartwright-jones1991, cartwright-chang1993}.} In some sense $\Lambda$ seems much more deeply related to the `inner' quantum mechanical structure of the system than the postulated common causes $\mathsf{C_{ij}^{ab}}$. (This observation is also supported by the fact that the $\mathsf{C_{ij}^{ab}}$ have been postulated in a purely classical context). Indeed, the postulated common causes $\mathsf{C_{ij}^{ab}}$ should not be thought of as hidden variables as such. For they are not aimed at completing in any sense the quantum description of the system.

We may assume furthermore that the causal factor $\Lambda$ is common to several (or even to all) correlations in the EPR experiment. It is important to note however that $\Lambda$, in contrast to the $\mathsf{C_{ij}^{ab}}$, will not in general screen-off the correlations. We shall explicitly require this in order to avoid problems concerning \emph{common}-common causes. That such problems may arise is clear, since if $\Lambda$ is a screening-off causal factor common to all the possible outcomes of the experiment, it surely is a \emph{common}-common cause of all the outcome correlations. And as pointed out a \emph{common}-common cause model cannot in general be guaranteed to exist. Moreover, in the EPR context, assuming \emph{common}-common causes leads quite straightforwardly to the Bell inequalities.

Similar reasons take us to require as well that the conjunction of $\Lambda$ with the measurement operation events, i.e.~$L_i \land R_j \land \Lambda$, be non-screening-off events. In particular, $L_i \land R_j \land \Lambda$ is assumed to be a causal factor common to all the correlations that involve these particular measurement settings. 

This suggests that the event $L_i \land R_j \land \Lambda$ \emph{contains} all the postulated common cause events $\mathsf{C_{ij}^{ab}}$ corresponding to the correlations displayed for these specific measurement settings. That is:
\begin{align}
L_i \land R_j \land \Lambda & \supseteq \mathsf{C_{ij}^{++}} \lor \mathsf{C_{ij}^{+-}} \lor \mathsf{C_{ij}^{-+}} \lor \mathsf{C_{ij}^{--}}.
\end{align}

One may however think that the above would entail that the $\mathsf{C_{ij}^{ab}}$ be \emph{common}-common causes. In order to avoid that and guarantee that $\mathsf{C_{ij}^{ab}}$ be \emph{individual}-common causes, we shall require that they be mutually exclusive, i.e.
\begin{align}
\mathsf{C_{ij}^{ab}} \land  \mathsf{C_{ij}^{a'b'}} & =  \emptyset,
\end{align}

\noindent where $a,b, a',b' = +, -$ and $ab \neq a'b'$. 

A remarkable feature of the resulting event structure is that our postulated common causes $\mathsf{C_{ij}^{ab}}$ satisfy a familiar condition very closely related to the derivation of the Bell inequalities, namely \emph{outcome independence} (OI):
\begin{equation}
\label{eq:oi}
p(L_i^a \land R_j^b \vert L_i \land R_j \land \mathsf{C_{ij}^{ab}} ) = p(L_i^a \vert
L_i \land R_j \land \mathsf{C_{ij}^{ab}} ) \cdot p(R_j^b \vert L_i \land R_j \land \mathsf{C_{ij}^{ab}} ),
\end{equation}  

This is indeed what expressions~\eqref{eq:model_RPCC_01} and~\eqref{eq:model_RPCC_02} encapsulate.\footnote{Elementary calculation shows that if~\eqref{eq:c_in_lirilambda} holds, expression~\eqref{eq:model_RPCC_01} entails equation~\eqref{eq:oi} above.} But note as well that, on the other hand, the event $\Lambda$ (and also $L_i \land R_j \land \Lambda$) will not in general satisfy such constraint. In particular, both $\Lambda$ and $L_i \land R_j \land \Lambda$ will in general violate \emph{outcome independence} due to the requirement that they do not screen-off the correlations ---recall this was explicitly required in order to avoid the usual problems regarding \emph{common}-common causes. Thus a model built on the premisses above would be able to accommodate the usual interpretation of the violations of the Bell inequalities, \emph{via} violations of \emph{outcome independence}, when $\Lambda$ is taken as the hidden variable.\footnote{This view hinges on the claim that, because of the spherical symmetry of the spin-singlet state, quantum mechanics violates \emph{outcome independence}, while it is compatible with \emph{parameter independence}.}

\subsection{Common cause measurement dependence is not conspiracy}
\label{ssec:ccdependence}
A second remarkable feature of a model as the outlined above is that, because of the dependence of the common causes on measurement, the $\mathsf{C_{ij}^{ab}}$ violate \emph{measurement independence}. That is:
\begin{align}
\label{eq:model_no-cons1}
p(\mathsf{C_{ij}^{ab}} \land L_i) & \neq p(\mathsf{C_{ij}^{ab}}) \cdot p(L_i), \\
p(\mathsf{C_{ij}^{ab}} \land R_j) & \neq p(\mathsf{C_{ij}^{ab}}) \cdot p(R_j).\label{eq:model_no-cons2}
\end{align}  

Recall that \emph{measurement independence} was a condition explicitly required in the standard view ---usually under the name of \emph{no-conspiracy}--- on the grounds that its violation would amount to some sort of `universal conspiracy'. The question then seems quite straightforward: in the light of expressions~\eqref{eq:model_no-cons1} and~\eqref{eq:model_no-cons2}, does the above constitute a `conspiratorial' model? Alternatively, is \emph{measurement independence} a reasonable assumption for common causes of EPR correlations? The answer to both these questions, I think, is negative. 

The charge of conspiracy may be avoided in different ways. Appealing to backwards in time causation is one of them, which I shall discuss only briefly in the next section.

Another possibility, more novel and interesting in my opinion, opens up if we look closely at the role of measurement in the experiment. In fact, one of the things I pointed out when discussing the significance of \emph{measurement independence} was that it seemed appropriate only if one assumed that the common causes take place prior to the measurement apparatus is set up (and therefore before measurement has been performed). It is only in those cases that probabilistic dependencies between the common causes and measurement settings, i.e.~violations of \emph{measurement independence}, may be sensibly interpreted as some sort of `world conspiracy'. However, there is nothing in the notion of common cause, nor in the very structure of the EPR experiment, that forces us to endorse the fact that the common cause must take place prior to measurement.

In fact, relations~\eqref{eq:rcc1}-\eqref{eq:rcc4} on page~\pageref{eq:rcc1} ---taken as a definition of (Reichenbachian) common cause--- do not include spatio-temporal information of any sort. This is mainly due to the fact that they involve event types, which are not defined in space time. It is only in virtue of them being collections of token events that we can refer to them spatio-temporally. And it is only under this view that common cause (type) events in expressions~\eqref{eq:rcc1}-\eqref{eq:rcc4} can be said to be located in the causal past of the correlated (type) events ---meaning that the corresponding token events follow the correct temporal sequence. In the particular case of the EPR experiment, we want to require that the postulated common causes and the corresponding outcome events be time-like (ensuring hence temporal priority of the common causes). But, once more, this needs not entail that such common causes be located prior to measurement operations.

\begin{figure}
\begin{center}
\includegraphics[scale=.53]{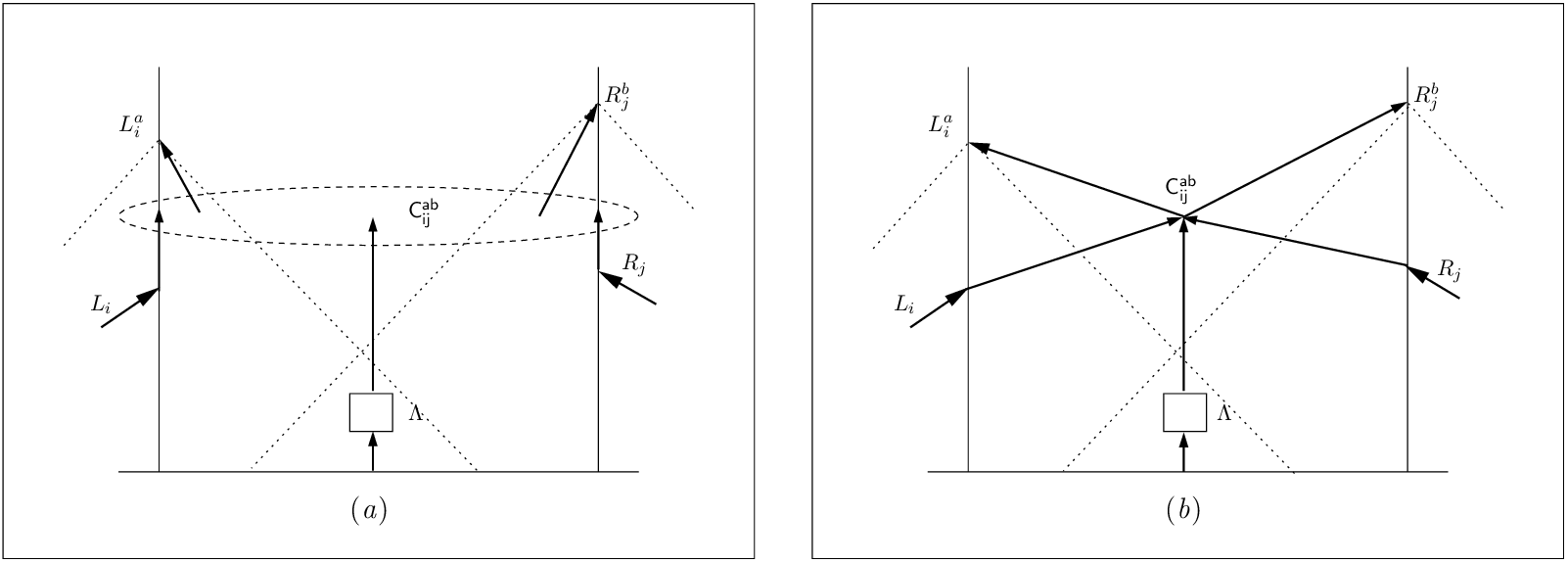}
\caption{Measurement dependent no-conspiratorial common causes may be thought of either (\emph{a}) as non-localised events spreading over some region of space-time, or (\emph{b}) as well localised common cause events reaching space-like separated regions.}\label{fig:model_space-time}
\end{center}
\end{figure}

So, we may perfectly allow instead for the common causes to take place after measurement operations have been performed ---and therefore after the measurement devices have been set-up as well.\footnote{The idea of common causes taking place after measurement can also be found in \cite{martel2008}. However, Martel's proposal does not have the same aims and scope than mine here. In particular, \citeauthor{martel2008} discusses specific issues as regards the philosophical status of the so-called \emph{causal Markov condition} ---a generalisation of RPCC---, and does not pay attenton to the consequences of the actual violation of \emph{measurement independence}, when it comes to locality, for instance.} In such a case, requiring  \emph{measurement independence} conditions to hold does not seem particularly appealing and, more importantly, there is nothing conspiratorial about them being violated. This is indeed the possibility the model above exploits. To illustrate how such measurement dependent no-conspiratorial common causes may look like two possibile conceptions are shown in Figure~\ref{fig:model_space-time}. Common causes may be thought of as non-localised events which spread over some region of space-time (Figure~\ref{fig:model_space-time}(\emph{a})). Or prehaps as well localised common cause events whose causal influences reach however space-like separated regions (Figure~\ref{fig:model_space-time}(\emph{b})). There is no need to discuss the details, virtues and possible problems of each of the two conceptions above. They are provided just as exemplifications for measurement dependent no-conspiratorial common causes and, of course, they need not be the only possibilities to conceive these.  

A possible objection to the view that common causes take place after measurement would be to note that since in this picture the notions `before' and `after' are frame dependent, the violation of \emph{measurement independence} will be interpretable as no-conspiratorial in some frames of reference only, i.e.~those in which the common cause events do in fact take place `after' measurement. In other frames of reference, the common cause events will take place `before' measurement and still violate \emph{measurement independence}, which would suggest, again, some kind of conspiratorial behaviour.

I see different possible responses to the objection above. One may argue in the first place that, while it is true that the notions `before' and `after' are frame dependent, they still have some significance, which one may take to be more relevant in some frames of reference than others. In particular, one may consider such notions more relevant, or perhaps more natural, if referred to the rest-frame of the laboratory. In this case, the model would still provide a possible non-conspiratorial common cause explanation of the EPR correlations (in the rest-frame of the laboratory). This view has however an obvious commitment (perhaps undesired by some) to the idea of the existence of a privileged frame of reference. 

As an alternative one could argue that frame dependence may not be as problematic for our purposes, after all. This, for two reasons, at least. On the one hand, note that the frame-dependence objection, as stated above, only shows that there is no common cause account of the EPR phenomena which violates \emph{measurement independence} and is not conspiratorial in \emph{all} frames of reference at once. What the objection does not challenge, though, is the idea of the existence of a (non-conspiratorial) common cause explanation of the EPR correlations in any particular frame of reference. Put it differently, we can still claim that, for any given (fixed) frame of reference, the model still provides a common cause account of the EPR correlations, which violates \emph{measurement independence} and yet is not conspiratorial. This seems to suggest that the model may be able to account for the EPR correlations in \emph{all} frames of reference, though the postulated common cause will be different for some of them. If this were the case then the problematic implications of the frame dependence of temporal terms could just be taken as an apparent feature of our (many) possible causal descriptions. 

This can only turn out to be a good solution though if one assumes that the \emph{common cause completability} theorems are capable of producing screening-off events with the right space-time features specific to each frame of reference. In particular, we need to make sure, for each frame of reference, that the screening-off events we postulate as common causes fall in the right temporal sequence with respect of the corresponding correlated events and measurement operations. However, \emph{common cause completability} by itself cannot ensure that this indeed is going to be the case. The solution above would require therefore that we impose some extra restrictions on the postulated screening-off events in order for these to be regarded as actual (common) causes. What such specific extra conditions will look like will depend, again, on the particular features of each frame of reference. We should want to know, for instance, whether their strength may vary in different frames, i.e.~whether the space-time features of certain frames of reference may set such strong space-time restrictions on the postulated screening-off events as to definitively rule out common cause explanations in them. In sum then, although it is not clear whether \emph{all} frames of reference can accommodate common cause explanations as suggested above, this option can neither be completely ruled out. 

Finally, we may want to stress the fact that what the model really aims at, is a \emph{causal} explanation of the correlations. As such, the key point to address here is what we understand by \emph{causal explanation}. In other words, the specific concept of causation one endorses will be crucial. Taking this into account, we shall then point out that there are several understandings of causation which do not incorporate ---nor they need to--- any reference to space-time features of the events involved. A causal account based on pure, so to speak, type events for instance, with no reference or reduction to token events ---which by the way would be completely compatible with the formulation of RPCC assumed here---, would do. Counterfactual accounts of causation constitute another way to understand causal relations without referring to a space-time structure.\footnote{I should note that the model is flexible enough not to commit to any particular account of causation even if, I have to admit, some idea of temporal order needs to be implemented if one is to challenge the requirement of \emph{measurement independence}.} The frame-dependence objection then looses much of its grip once such views are endorsed.

\section{Measurement dependence and locality}
\label{sec:mdependence}
As a direct consequence of the violation of \emph{measurement independence} the model does not face the charge of Bell's theorem. This is so simply because \emph{measurement independence} is necessary for the derivation of the Bell inequalities. This is certainly good, for it means that the model allows for the violation of the Bell inequalities and is therefore perfectly compatible with the (empirically confirmed) quantum mechanical predictions for the EPR correlations. However, the fact that the model's common causes are explicitly \emph{measurement dependent} seems to have implications as regards locality issues. 

The first thing to note in this respect is that it is not the fact that the postulated common causes are measurement dependent in itself that opens the door to non-locality. The crucial issue, together with the violation of \emph{measurement independence}, is the specific temporal order assumed. More specifically, it is the actual requirement that measurement dependent common causes \emph{do not} take place prior to measurement that seems to lead to conflicts with locality. This becomes clear if we take a look at models featuring backwards in time causal influences, where the idea of locality is retained even if measurement operations and common causes (located in their remote past) fail to be statistically independent.

The possibility of backwards in time influences is not, by any means, new. It was explored for instance in a common cause model proposed by \cite{price1994, price1996a}.\footnote{As \cite{suarez2007} points out, although the model was initially presented as a common cause model (containing backwards in time causal influences), \citeauthor{price1996b} seemed later to retract from interpreting it as causal. In particular, \cite{price1996b} seems to suggest that the backwards in time influences of the model be of no causal origin. \cite{suarez2007} also provides an explicit causal interpretation of \citeauthor{price1994}'s model, which I endorse here.} \citeauthor{price1994}'s model assumes that the common cause events take place in the overlap of the backward light-cones of the correlated EPR outcomes. This guarantees that they act locally to produce the corresponding outcomes. The common causes in the model, however, are causally influenced (backwards in time) by the measuring operations in both wings. Thus the model explicitly violates \emph{measurement independence}. In sum, in \citeauthor{price1994}'s model backwards in time influences are what make it possible for the postulated (measurement dependent) common causes to operate locally.\footnote{Note that the probabilistic event structure of \citeauthor{price1994}'s model and my own is exactly the same. Only, the interpretation of the events is different.}

As a matter of fact then, when faced with violations of \emph{measurement independence}, we seem bound to choose between two type of models, depending on which time ordering is assumed for the events. A first possibility would be to assume the postulated common causes to be located in the remote (enough) past of the corresponding outcomes. In this case they will operate locally to produce these outcomes, but will also need to involve some backwards in time causal influences (from the experimenter's measurement choices or operations on the common causes). I do not see backwards causation as particularly appealing (mainly for intuitive reasons), though. As an alternative, we may want to locate the postulated common causes in the future of the measuring operations, as suggested above. No backwards in time influences are needed in this case but it remains open whether such a picture involves some sort of non-locality. The question, more specifically, is whether locality can be retained in a model which violates \emph{measurement independence} but has no backwards in time causal influences. My own personal intuitions point to a negative answer to this question but it seems worthwhile having a closer look to the issue. 

As a first reaction, for instance, it seems worth noting that in developing the model we have only paid attention to the significance and adequacy of \emph{measurement independence}. But recall that \emph{measurement independence} (or the equivalent \emph{no-conspiracy}) conditions are just some of the extra restrictions imposed on the idea of Reichenbachian common cause, alongside other further conditions, some of them associated to locality requirements. What this means is that, by construction, our model is supposed to have retained some notion of locality within its event structure. In other words, while the model explicitly violates \emph{measurement independence}, it may well be the case that it satisfies the specific locality assumptions we had imposed onto the postulated common causes. It would be interesting to see, in case this is so, in what sense we could make sense of the model being local (as required by some specific locality conditions), despite the failure of \emph{measurement independence} to hold. Addressing such issues would require formulating the specific locality conditions and investigating their relation to \emph{measurement independence}, and deserve a paper on its own.

Our hopes to retain locality ---at least in some standard notion--- do not seem too promising, though. For violations of \emph{measurement independence} might well be related to violations of \emph{parameter independence}, which would in turn imply a violation of Bell's \emph{factorizability}~\citep{bell1975}. It is not completely clear whether this is really so and I shall leave it as a conjecture at this point:

\begin{conject}
If \emph{measurement independence} is violated then \emph{parameter independence} is also violated.\label{conject:mipi}
\end{conject}

As I just pointed out, it is not clear whether Conjecture~\ref{conject:mipi} is true or false but one can find support for the claim it makes in the following informal intuitive argument. 

Note first that a violation of \emph{measurement independence} entails, as we have seen in the previous section, that the common cause depends on \emph{both} measurement settings. That is:
\begin{align}
p(\mathsf{C_{ij}^{ab}} \land L_i) & \neq p(\mathsf{C_{ij}^{ab}}) \cdot p(L_i),  \\
p(\mathsf{C_{ij}^{ab}} \land R_j) & \neq p(\mathsf{C_{ij}^{ab}}) \cdot p(R_j),
\end{align}  

\noindent which entails that
\begin{align}
p(\mathsf{C_{ij}^{ab}} \land L_i \land R_j) & \neq p(\mathsf{C_{ij}^{ab}}) \cdot p(L_i \land R_j),
\end{align}  
\noindent as long as we assume $p(L_i)$ and $p(R_j)$ probabilistically independent.

This in turn seems to suggest that the outcomes will also depend on \emph{both} measurement settings, since they obviously depend on the common cause, i.e.
\begin{align}
p(L_i^a \vert L_i \land R_j \land \mathsf{C_{ij}^{ab}}) & \neq p(L_i^a \vert L_i \land  \mathsf{C_{ij}^{ab}}),  \\
p(R_j^b \vert L_i \land R_j \land \mathsf{C_{ij}^{ab}}) & \neq p(R_j^b \vert R_j \land  \mathsf{C_{ij}^{ab}}).
\end{align}  

This expression is nothing more than the violation of \emph{parameter independence}. 

I would like to stress that whether Conjecture~\ref{conject:mipi} turns out to be true or not is not at all crucial for the claim in the previous sections that the model is not committed to the charge of Bell's theorem to follow. For, as I have pointed out, \emph{measurement independence} is already \emph{necessary} for the Bell inequalities to be derived. That is, the Bell inequalities cannot be derived if \emph{measurement independence} is violated, regardless of whether or not \emph{parameter independence} is violated as well.\footnote{Of course, if the my conjecture turns out to be true, \emph{parameter independence} would be violated in all cases where \emph{measurement independence} would fail to hold. In other words, we could not have a model satisfying \emph{parameter independence} while \emph{measurement independence} was violated in it. But, on the other hand, because of the logical structure of the conjecture we could have models satisfying \emph{measurement independence}, where \emph{parameter independence} was nevertheless violated. This is the case in Bohm's quantum mechanics, for instance.}

Again, the main thrust and significance of the claim in the conjecture is related, in my opinion, to its consequences as regards the local/non-local character of the model. In fact, did Conjecture~\ref{conject:mipi} turn out to be true, it would seem to provide the grounds to claim that the model is non-local. More specifically, since \emph{parameter independence} is necessary for Bell's \emph{factorizability} then, by Conjecture~\ref{conject:mipi}, a violation of \emph{measurement independence} would also entail that \emph{factorizability} is violated. This is generally taken to be a sign of non-local behaviour.\footnote{This diagnosis is not free of controversies, however. Several authors in fact cast doubts as to whether \emph{factorizability} indeed reflects the idea of physical locality ---especially if locality is merely associated with the requirement that there not be superluminal signalling between the two wings of the EPR experiment. See for instance, \cite{wessels1985, fine1986} or \cite{maudlin1994}.} Furthermore, and also in support of the idea that the model might turn out to be non-local after all, were Conjecture~\ref{conject:mipi} be confirmed, one could attempt to draw some parallelisms between the model and Bohm's quantum mechanics~\citep{bohm1952}, which is explicitly non-local. For both of them would be seen to violate \emph{parameter independence}, while conforming to \emph{outcome independence}. 

However, the possible resemblances between the model here and Bohm's quantum mechanics are limited and refer, I would say, exclusively to the very fact that in both cases \emph{parameter independence} fails and \emph{outcome independence} is satisfied. There is no resemblance at all, for instance, as to \emph{how} is \emph{parameter independence} violated in each case. Put it another way, while it is true that both the model here and Bohm's theory violate \emph{parameter independence}, it is not less true that they do it for different reasons. As we have seen, the model's violation of \emph{parameter independence} comes from a failure of \emph{measurement independence} (always assuming that Conjecture~\ref{conject:mipi} is true, of course). But this is not so in Bohm's theory, which indeed satisfies \emph{measurement independence}.\footnote{Recall that because of its logic asymmetry, Conjecture~\ref{conject:mipi} can accommodate violations of \emph{parameter independence} in cases \emph{measurement independence} holds.} This means, in terms of locality/non-locality, that the non-local character that the model displays has a different source ---it may even be fundamentally different in itself--- than that in Bohm's quantum mechanics. In particular, when it comes to locality issues, the model tells us something about EPR that Bohmian mechanics does not, namely that measurement operations are causally relevant to the outcomes in a specific way ---they are constitutive of the common causes. Thus what the model provides, as compared with Bohmian mechanics, is a different approach to causality and locality.

One may argue however that also Bohm's theory is capable of a causal explanation of EPR (even a better one perhaps than that provided here). So why should we go for our common cause model instead?

We should note first, and before addressing this possible objection, that the aim and purpose of the model here are not comparable in many senses to that of Bohmian mechanics. In particular, the model has been suggested with the only aim to provide a causal explanation of the EPR phenomena and, of course, it does not provide ---nor it was the underlying motivation at all--- the precise elements for a detailed quantum mechanical description of EPR as such.\footnote{Not even at the ontological level can they be compared, I think. For while Bohm's quantum mechanics provides a definite ontological picture for quantum mechanics, the ontology associated to the common causes in the model is not to be taken as the quantum ontology ---again, it is key to bear in mind that the common causes are not to be seen as hidden variables as such. This is not to say, of course, that the possible ontologies we might want to provide the model with will not `inherit' somehow, or reflect, some quantum features, such as non-locality.} Thus, the question above should not be taken as a question about whether one should take stands either for the model I have proposed here or for Bohmian mechanics. In my opinion, each option should be judged by its own merits.

Having said that, one of the reasons the model may be of use and interest can be found precisely in our discussion on non-locality above. As I said, the model seems to convey a different notion of non-locality than that present in Bohmian mechanics. This, it seems to me, is already interesting in itself, in that it provides an approach to the issue of locality/non-locality from a different perspective. Furthermore, the causal account provided by the model is also different in character. As we have seen, the causal explanation provided by the model has its origins in the very phenomenology of the EPR experiment. Let me stress again that the common cause model here does not constitute in any sense an attempt to complete the quantum formalism ---again, the fact that common cause are not to be understood as the usual hidden variables, reinforces this point--- but just to provide a causal explanation of the EPR phenomena. And the assumptions the model builds on mainly draw from the \emph{classical} intuitions about the notion of causation, as reflected in Reichenbach's Principle of the Common Cause. In that sense thus, the common cause model here is not comparable at all with Bohm's quantum theory.

What the model really constitutes then is a testing ground for some of our most rooted intuitions about notions such as causality and locality in the context of quantum mechanics. For instance, and going back to the issue of the specific locality conditions that the model may well be thought to satisfy, we saw that there might be some reasons to consider the model's common causes as local, even if \emph{measurement independence} is not satisfied. The question seems to be then, whether the notion of locality which one can find associated to \emph{measurement independence} conditions differs in any fundamental sense from the more specific notions of locality used in the derivation of the Bell inequalities. The answers to questions such as that require a close evaluation of the different notions of physical locality available, including that which is arguably present in Bell's \emph{factorizability} condition, as well as a revision of our most immediate intuitions about the nature of causation.



\begin{thebibliography}{Name YY}
\bibliographystyle{spphys}
\bibitem[Bell(1964)]{bell1964} {B}ell, J. S. (1964). On the {E}instein-{P}odolsky-{R}osen paradox. \emph{Physics},  \textbf{1}, 195--200. Reprinted in J. S. {B}ell, \emph{Speakable and unspeakable in quantum mechanics} (pp.~14--21). Cambridge University Press, 1987.

\bibitem[Bell(1975)]{bell1975} {B}ell, J. S. (1975). The theory of local beables. TH-2053-CERN, presented at the Sixth GIFT Seminar, June 1975. Reproduced in \emph{Epistemological Letters}, March 1976, and reprinted in J. S. {B}ell, \emph{Speakable and unspeakable in quantum mechanics}. Cambridge University Press, 1987, pp.~52--62.

\bibitem[Berkovitz(2002)]{berkovitz2002} Berkovitz, J. (2002). On causal loops in the quantum realm. In T. Placek and J. Butterfield (Eds.), \emph{Non-locality and modality: Proceedings of the NATO advanced research workshop on modality, probability and {B}ell's theorems}. Kluwer, 2002, pp.~235--257.

\bibitem[Bohm(1952)]{bohm1952} Bohm, D. (1952). A Suggested interpretation of the quantum theory in terms of `hidden' variables, I and II. \emph{Physical Review}, \textbf{85}, 166--193.

\bibitem[Butterfield(1989)]{butterfield1989} Butterfield, J. (1989). A space-time approach to the {B}ell inequality. In J. Cushing and E. McMullin (Eds.), \emph{Philosophical consequences of quantum theory}. University of Notre Dame Press, 1989, pp.~114--144.

\bibitem[Butterfield(2007)]{butterfield2007} Butterfield, J. (2007). Stochastic {E}instein's locality revisited. \emph{British Journal for the Philosophy of Science}, \textbf{58}, 805--867.

\bibitem[Cartwright(1987)]{cartwright1987} Cartwright, N. (1987). How to tell a common cause: Generalizations of the conjunctive fork criterion. In J. H. Fetzer (Ed.), \emph{Probability and causality}. Reidel Pub. Co., 1987, pp.~181--188.

\bibitem[Cartwright \& Jones(1991)]{cartwright-jones1991} Cartwright, N. \& Jones, M. (1991). How to hunt quantum causes. \emph{Erkenntnis}, \textbf{35}, 205--231.

\bibitem[Chang \& Cartwright(1993)]{cartwright-chang1993} Chang, H. \& Cartwright, N. (1993). Causality and realism in the {EPR} experiment. \emph{Erkenntnis}, \textbf{38}, 169--190.

\bibitem[Einstein et al.(1935)]{epr1935} Einstein, A., Podolsky, B. \& Rosen, N. (1935). Can quantum-mechanical description of physical reality be considered complete?.  \emph{Physical Review}, \textbf{47}, 777--780.

\bibitem[Fine(1981)]{fine1981} Fine, A. (1981). Correlations and physical locality. In P. Asquith and R. Giere (Eds.), \emph{Proceedings of the 1980 biennial meeting of the Philosophy of Science Association, Vol. 2}. Lansing, 1981, pp.~535--556.

\bibitem[Fine(1986)]{fine1986} Fine, A. (1986).  \emph{The shaky game. {E}instein realism and the quantum theory}. The University of Chicago Press (2nd edition, 1996).

\bibitem[Gra{\ss}hoff et al.(2005)]{grasshoff2005} Gra{\ss}hoff, G., Portman, S. \& W{\"{u}}thrich, A. (2005). Minimal assumption derivation of a {B}ell-type inequality. \emph{The British Journal for the Philosophy of Science}, \textbf{56}, 663--680.

\bibitem[Healey(1992)]{healey1992} Healey, R. (1992). Chasing quantum causes: How wild is the goose?. \emph{Philosophical Topics}, \textbf{20}, 181--204.

\bibitem[Henson(2005)]{henson2005} Henson, J. (2005). Comparing causality principles. \emph{Studies in the History and Philosophy of  Modern Physics}, \textbf{36}, 519--543.

\bibitem[Hofer-Szab{\'{o}} et al.(1999)]{hofer-szabo1999} Hofer-Szab{\'{o}}, G., R{\'{e}}dei, M. \& Szab{\'{o}},  L. E. (1999). On {R}eichenbach's common cause principle and {R}eichenbach's notion of common cause. \emph{The British Journal for the Philosophy of Science}, \textbf{50}, 377--399.

\bibitem[Hofer-Szab{\'{o}} et al.(2000)]{hofer-szabo2000a}  Hofer-Szab{\'{o}}, G., R{\'{e}}dei, M. \& Szab{\'{o}},  L. E. (2000). Common cause completability of classical and quantum probability spaces. \emph{International Journal of Theoretical Physics}, \textbf{39}, 913--919.

\bibitem[Hofer-Szab{\'{o}} et al.(2002)]{hofer-szabo2002} Hofer-Szab{\'{o}}, G., R{\'{e}}dei, M. \& Szab{\'{o}},  L. E. (2002). Common causes are not common-common causes. \emph{Philosophy of Science}, \textbf{69}, 623--636.


\bibitem[Martel (2008)]{martel2008} Martel, I. (2008). The Principle of the Common Cause, the Causal Markov Condition, and Quantum Mechanics: Comments on Cartwright. In S. Hartmann, C. Hoefer, and L. Bovens (Eds.), \emph{Nancy Cartwright's Philosophy of Science}. Routledge, 2008, pp.~242--262.

\bibitem[Maudlin(1994)]{maudlin1994} Maudlin, T. (1994). \emph{Quantum non-locality and relativity}. Blackwell Publishing (2nd edition, 2002).

\bibitem[Price(1994)]{price1994} Price, H. (1994). A neglected route to realism about quantum mechanics. \emph{Mind}, \textbf{103}, 303--336.

\bibitem[Price(1996a)]{price1996a} Price, H. (1996a). Locality, independence and the pro-liberty Bell. \url{http://arxiv.org/abs/quant-ph/9602020}.

\bibitem[Price(1996b)]{price1996b} Price, H. (1996b). \emph{Time's arrow and {A}rchimedes' point}. Oxford University Press, 1996.

\bibitem[Reichenbach(1956)]{reichenbach1956} Reichenbach, H. (1956). \emph{The direction of time\textup{; Edited by Maria Reichenbach}}. Unabridged Dover, 1999 (republication of the original University of California Press, 1956).

\bibitem[San Pedro \& Su{\'a}rez(2009)]{sanpedro-suarez2009} San Pedro, I. \& Su{\'a}rez, M. (2009). Reichenbach's principle of the common cause and indeterminism: A review. In J. L. Gonz{\'a}lez Recio (Ed.), \emph{Philosophical essays in physics and biology}. Georg Olms, 2009, pp.~223--250.

\bibitem[Su{\'a}rez(2000)]{suarez2000} Su{\'a}rez, M. (2000). The many faces of non-locality: Dickson on the quantum correlations. \emph{British Journal for the Philosophy of Science}, \textbf{51}, 882--92.

\bibitem[Su{\'a}rez(2007)]{suarez2007} Su{\'a}rez, M. (2007). Causal inference in quantum mechanics: A reassessment. In F. Russo and J. Williamson (Eds.) \emph{Causality and Probability in the Sciences} (pp.~65--106). London College.

\bibitem[Szab{\'o}(2000)]{szabo2000} Szab{\'o}, L. E. (2000). On an attempt to resolve the {EPR}-{B}ell paradox via {R}eichenbachian concept of common cause. \emph{International Journal of Theoretical Physics}, \textbf{39}, 911--926.

\bibitem[Szab{\'o}(2008)]{szabo2008} Szab{\'o}, L. E. (2008). The {E}instein-{P}odolsky-{R}osen argument and the {B}ell inequalities. \emph{The Internet Encyclopaedia of Philosophy}, \url{http://www.iep.utm.edu/}.

\bibitem[van Fraassen(1982)]{vanfraassen1982a} van Fraassen, B C. (1982). The charybdis of realism: Epistemological implications of {B}ell's Inequality. \emph{Synthese}, \textbf{52}, 25--38. Reprinted with corrections in J. Cushing and E. McMullin (Eds.), \emph{Philosophical consequences of quantum theory}. University of Notre Dame Press, 1989, pp.~97--113.

\bibitem[Wessels(1985)]{wessels1985} Wessels, L. (1985). Locality, factorability and the {B}ell inequalities. \emph{No{\^u}s}, \textbf{19}, 481--519.

\end{thebibliography}

\end{document}